\documentclass[preprint,12pt]{elsarticle}

\usepackage{amsmath}
\usepackage{amssymb}
\usepackage{graphicx}
\usepackage{bm}
\usepackage{setspace}
\usepackage{subfig}
\usepackage{pdflscape}
\usepackage{lscape}
\usepackage{color}

\newcommand{\non}{\nonumber}

\newcommand{\be}{\begin{equation} }
\newcommand{\ee}{\end{equation}}
\newcommand{\bea}{\begin{eqnarray} } 
\newcommand{\eea}{\end{eqnarray} }
\newcommand{\sg}{\sigma} 
\newcommand{\gm}{\gamma}
\newcommand{\Gm}{\Gamma}
\newcommand{\hb}{\hbar}

\newcommand{\og}{\omega}

\newcommand{\lb}{\lambda}

\newcommand{\tbf}{\textbf}
\newcommand{\bds}{\boldsymbol}
\newcommand{\kp}{\kappa}

\newcommand{\ep}{\epsilon}

\newcommand{\comment}[1]{}
\newcommand{\alld}{\allowdisplaybreaks}
\newcommand{\ta}{\theta}
\newcommand{\zt}{\zeta}

\newcommand{\im}{\imath}

\newcommand{\vta}{\vartheta}

\newcommand{\Lb}{\Lambda}

\newcommand{\lgl}{\langle}
\newcommand{\rgl}{\rangle}

\newcommand{\Tr}{{\rm Tr}}
\newcommand{\Rl}{{\rm Re}}

\journal{Physics Letters A}

\begin{document}

\begin{frontmatter}

\title{Dynamics and non-equilibrium steady state in a system of coupled harmonic oscillators}

 \author[sa]{Anne Ghesqui\`{e}re\fnref{fn1}}
 \ead{Anne.Ghesquiere@nithep.ac.za}
 \author[sa]{Ilya Sinayskiy}
 \ead{sinayskiy@ukzn.ac.za}
 \author[sa]{Francesco Petruccione}
 \ead{petruccione@ukzn.ac.za}
\address[sa]{School of Chemistry and Physics and National Institute for Theoretical Physics, Westville Campus, UKZN, Private Bag X54001, Durban, 4000, South Africa}
\fntext[fn1]{Current address: Department of Mathematical Physics, National University of Ireland, Maynooth, Co. Kildare, Ireland}

\date{\today}

\begin{abstract}
A system of two coupled oscillators, each of them  coupled to an independent reservoir, is analysed. The analytical solution of the non-rotating wave master equation is obtained in the high-temperature and weak coupling limits.  No thermal entanglement is found in the high-temperature limit. In the weak coupling limit the system converges to an entangled non-equilibrium steady state. A critical temperature for the appearance of quantum correlations is found.
\end{abstract}

\end{frontmatter}

\section{Introduction}
Dissipation, as induced by the coupling of one or several reservoirs to a quantum system, is generally understood to play a crucial role in the evolution of the state of the system \cite{Petru:2002}. Its effect on the evolution of the entanglement within the system has been widely studied for a variety of systems. For instance, the effects of baths at different temperatures at the end of a spin chain were studied in Refs. \cite{Ilya:2008, Ilya:2011}. For this situation it was found that the system converges to the non-equilibrium steady state. The dynamics of entanglement was also studied in the context of two non-identical oscillators coupled to an environment \cite{Zambrini:2010} and the effect of a diversity in frequencies of the oscillators was investigated. The quantum discord, a measure of quantum correlations, was studied within a system of two oscillators coupled to the same heat bath \cite{Freitas:2012}. Furthermore, the destruction of the entanglement through the interaction with a reservoir was coined entanglement sudden death by Eberly et al. \cite{PE:2001, YE:2003, YE:2004, YE:2006}. However, it is also well-known that the presence of an environment may restore or even create entanglement. In this way, Ficek and Tan\'as have shown that initially separable qubits may become entangled after a certain time, through spontaneous emission \cite{Ficek:2008}. Krauter et al. \cite{Krauter:2011} have examined a pair of two-level systems interacting with a common $y$-polarised laser field. The atoms are then placed in a $x$-polarised magnetic field. The coupling creates an entangling mechanism which allowed them to generate entanglement and to maintain it in the steady state. Further examples of entanglement revivals have been reported \cite{Ficek:2006, Lopez:2012}.

It is typical to have one small quantum system interacting with a dissipative environment. The dissipative effects are understood to bring the quantum system to equilibrium. Here, a system of two particles, initially entangled and with a Gaussian wavefunction is studied. Each particle is coupled to its own thermal reservoir and each reservoir is assumed to be at its own temperature. Many such situations are studied through the stationary behaviour of the entanglement \cite{Briegel:2006, Quiroga:2007}. It was shown that if the particles are interacting strictly harmonically, there is no steady state within the considered assumptions \cite{sakuya:2012}. It was found that to have entanglement at high temperature the interaction must be strong. The emergence of thermal entanglement is associated with the approach of the system to equilibrium. This phenomenon has been termed thermalisation \cite{Ilya:2008, Ilya:2011} and has been studied with a variety of witnesses. For instance, the energy was used to examine the thermalisation of the system \cite{Marchiori:2012}. The concept of thermal entanglement was first pointed out by Braun \cite{Braun:2002}, who shows that entanglement may be created between two qubits, which are interacting with a common environment, but not with each other. 

However, not every physical system is entangled in the steady state. In this case one can use the entropy as the witness of thermalisation. The von Neumann entropy is chosen here, since it has a simple formulation in terms of the covariance matrix. Indeed, we keep to continuous variables and more particularly, to Gaussian states, which allow for an elegant mathematical treatment \cite{Eisert:2003, Eisert:2004, Anders:2003} and for the explicit computation of the entanglement through the logarithmic negativity \cite{Werner:2002}.

The article is structured as follows. In Section~\ref{time_evolution}, a pre-Lindblad master equation is presented for two different limiting cases. The first case is the quantum Brownian motion limit (\cite{sakuya:2009, sakuya:2010}) for a system of two oscillators with a linear interaction. The second case is the weak coupling limit \cite{O'C:2003}. An analytical solution in both cases is presented. A fully general master equation was derived recently \cite{Martinez:2012}. However, the master equation used here is appropriate for our purpose. In Section~\ref{entgt-dyn}, the evolution of the entanglement between the oscillators is studied and properties of thermal entanglement are investigated. An approach to equilibrium is studied in Section~\ref{entropy}. We conclude in Section~\ref{conclusion}. Explicit mathematical expressions of the solution are presented in \ref{App_solution}. The steady state solution is presented in \ref{steady_state}.

\section{Time evolution of two interacting harmonic oscillators \label{time_evolution}}
The dynamics of the entanglement in a bipartite system coupled to an environment is greatly influenced by allowing an harmonic interaction between the system's particles \cite{sakuya:2010, Ficek:2006, Ficek:2008}. We study a system of two particles of equal mass, each one coupled to its own heat bath; they have coordinates $x_1$ and $x_2$, momenta $p_1$ and $p_2$; $\og_0$ denotes the frequency of their oscillations. We examine the case where the particles are interacting linearly. The overall Hamiltonian reads
\begin{align}
 H=& \frac{p_1  ^2}{2m} + \frac{p_2 ^2}{2m} + \frac{m \og_0 ^2}{2} (x_1 ^2 + x_2 ^2) + \kp x_1 x_2 \non
 \\ &+ \sum_j \left\{\frac{p_j ^2}{2 m_j} + \frac{m_j \og_j ^2}{2} (q_j - x_1)^2 \right\} + \sum_k  \left\{\frac{p_k ^2}{2 m_k} + \frac{m_k \og_k ^2}{2} (q_k - x_2)^2 \right\}  ,\end{align}
where $q_{j,k}$ and $p_{j,k}$ are the positions and momenta of the oscillators in the bath. For our study, we take all masses to be equal. The frequencies of the bath oscillators are denoted by $\og_i$. The initial state is chosen to be the Gaussian state \cite{FO'C:2008, FO'C:2010}
\begin{equation}
\Psi(x_1, x_2) = \sqrt{\frac{1}{2 \pi s d}} e^{-\frac{(x_1 - x_2 ) ^2}{4 s ^2}} e^{-\frac{(x_1 + x_2)
^2}{16d ^2}} \, ,\end{equation}
where $s$ and $d$ denote the distance between the particles and the width of the center-of-mass system, respectively. We assume a position coupling between the reservoirs and the particles; the Non-Rotating-Wave master equation in the quantum Brownian motion limit is written as
\begin{align}\label{eq:HiT}
 {\dot \rho} = - \frac{\im}{\hb} \left[ H_s , \rho \right] -& \frac{\im \gm_1}{2m\hb} \left[ x_1, \left[ p_1 , \rho \right]_+ \right] - \frac{\gm_1 k T_1}{\hb ^2} \left[ x_1, \left[ x_1, \rho \right] \right] \non 
\\ -& \frac{\im \gm_2}{2m\hb} \left[ x_2, \left[ p_2 , \rho \right]_+ \right] - \frac{\gm_2 k T_2}{\hb ^2} \left[ x_2, \left[ x_2, \rho \right] \right] \, .
\end{align}
This master equation describes the dynamics of the quantum harmonic oscillator in the high temperature regime ($T/\omega \gg 1$), without additional constraints on the strength of the system-environment interaction. For the description of the dynamics of the system at low temperatures we can use the weak coupling limit ($\gamma\ll \omega, m, k$) and in this case the Non-Rotating-Wave master equation in weak coupling limit reads \cite{O'C:2003},
\begin{align}\label{eq:LoT}
 {\dot \rho} = - \frac{\im}{\hb} \left[ H_s , \rho \right] -& \frac{\im \gm_1}{2m\hb} \left[ x_1, \left[ p_1 , \rho \right]_+ \right] - \frac{\gm_1 \omega}{2\hb} \coth{\frac{\hb\omega}{2 k T_1}}\left[ x_1, \left[ x_1, \rho \right] \right] \non 
\\ -& \frac{\im \gm_2}{2m\hb} \left[ x_2, \left[ p_2 , \rho \right]_+ \right] - \frac{\gm_2 \omega}{2\hb} \coth{\frac{\hb\omega}{2 k T_2}} \left[ x_2, \left[ x_2, \rho \right] \right] \, .
\end{align}
It is clear that the master equation (\ref{eq:LoT}) in the high temperature case will take the form of Eq. (\ref{eq:HiT}), for $T/\omega\gg1$ a hyperbolic cotangent can be approximated, by $\coth{\frac{\hb\omega}{2 k T}} \approx \frac{2 k T}{\hb\omega}$. From the mathematical point of view both master equations have the same form. In this article we demonstrate explicitly the analytical solution for the quantum Brownian particle Eq. (\ref{eq:HiT}). The explicit solution in the weak coupling limit Eq. (\ref{eq:LoT}) can be obtained by the formal substitution in the solution for the quantum Brownian motion limit Eq. (\ref{eq:HiT}) of $kT_i\rightarrow \frac{\omega}{2}\coth{\frac{\hb \omega}{2 k T_i}}$.

In order to solve the master equation (\ref{eq:HiT}) we write the density matrix in position representation, $\rho(x_1, x_2 ; y_1, y_2)$, and get 
\begin{align} \label{master_equation}
\frac{\partial \rho}{\partial t} =&  \frac{\im \hb}{2 m} \left( \frac{\partial ^2}{\partial x_1 ^2} - \frac{\partial ^2}{\partial y_1 ^2} + \frac{\partial ^2}{\partial x_2 ^2} -\frac{\partial ^2}{\partial y_2 ^2 } \right) \rho \non 
\\ &- \frac{\im m \og_0 ^2}{2\hb} \left( x_1 ^2 + x_2 ^2 - y_1 ^2 - y_2 ^2 \right) \rho - \frac{\im \kp}{\hb} (x_1 x_2 - y_1 y_2) \non 
\\ &- \frac{\gm_1}{2 m} \left( x_1 - y_1 \right) \left( \frac{\partial }{\partial x_1} - \frac{\partial}{\partial y_1} \right) \rho - \frac{\gm_1 k T_1}{\hb ^2} (x_1 - y_1)^2 \rho \non 
\\ &- \frac{\gm_2}{2 m} \left( x_2 - y_2 \right) \left( \frac{\partial }{\partial x_2} - \frac{\partial}{\partial y_2} \right) \rho - \frac{\gm_2 k T_2}{\hb ^2} (x_2 - y_2)^2 \rho \, .
\end{align}
The derivation of the solution is highlighted in \ref{App_solution}, following Refs. \cite{Gard:2000, sakuya:2009, sakuya:2010}. It begins by the change of variables $x = u + \hb z$, $y = u - \hb z$ and $\rho(\tbf{x},\tbf{y},t) \rightarrow P(\tbf{u}, \tbf{z}, t)$ is performed. Then a Fourier transform is applied
\be {\tilde P}(\tbf{q}, \tbf{z}, t) = \int du_1 \, du_2 \, P(\tbf{u}, \tbf{z}, t) e^{- \im q_1 u_1 - \im q_2 u_2}. \non \ee
For the missing steps, the reader is referred to \ref{App_solution}. The final solution yields
{\alld
\begin{align}
\tilde{P}(\tbf{q},\tbf{z},t) = & \exp\left(- \mathcal{A}_1 q_1 ^2 - \mathcal{A}_2 q_2 ^2 - \mathcal{B}_1 z_1 ^2 - \mathcal{B}_2 z_2 ^2 - \mathcal{E} q_1 q_2 - \mathcal{D} z_1 z_2\right) \non
\\ & \times \exp \left( - \mathcal{C}_{11} z_1 q_1 - \mathcal{C}_{22} z_2 q_2 - \mathcal{C}_{12} z_1 q_2 - \mathcal{C}_{21} z_2 q_1 \right)\, .
\end{align}}
The complete expressions for the coefficients may be found in \ref{App_solution}.
To study the entanglement, we use the logarithmic negativity, whilst to show the approach to equilibrium, we choose to look at the entropy because both of these measures may be expressed in terms of the covariance matrix $\Gm$. We note here that since we study Gaussian states, we may determine the covariance matrix in terms of second moments only, $\Gm_{jk} = 2 \Rl \Tr \left[ \rho \hat{R}_j \hat{R}_k\right]$.The $\hat{R}$'s are elements of the vector $[\hat{x}_1, \hat{p}_1, \hat{x}_2, \hat{p}_2]$. The covariance matrix $\Gm$ can be written as: 
\begin{align} \Gm = \left[
\begin{array}{cccc} 4 \mathcal{A}_1 & - \mathcal{C}_{11} & 2 \mathcal{E} & - \mathcal{C}_{21} \\ - \mathcal{C}_{11} & \mathcal{B}_1 & - \mathcal{C}_{12} & \mathcal{D}/2 \\ 2 \mathcal{E} & - \mathcal{C}_{12} & 4 \mathcal{A}_2 & - \mathcal{C}_{22} \\ - \mathcal{C}_{21} & \mathcal{D}/2 & - \mathcal{C}_{22} & \mathcal{B}_2 
\end{array}
\right] .\end{align}
The logarithmic negativity is obtained in terms of the symplectic eigenvalues of the partially transposed covariance matrix $\Gm^{T_1}$. This accounts here to sending $\hat{p}_1$ to $-\hat{p}_1$ \cite{Werner:2002}. The symplectic eigenvalues of $\Gm^{T_1}$ are defined as the positive square root of the eigenvalues of the matrix $- \sg \Gm^{T_1} \sg \Gm^{T_1}$, where $\sg = \bigoplus^{N}_{j=1}\left( \begin{array}{cc}0 & 1 \\ -1 & 0 \end{array}\right)$. We note here that two eigenvalues are degenerate, such that the logarithmic negativity then is
\be \mathcal{L_{\mathcal{N}}}(\rho) = -2\left( \log_2\left(\min(1,\vert \lb_{1,2} ^T \vert)\right)+\log_2\left(\min(1,\vert \lb_{3,4} ^T \vert)\right) \right). \non \ee 
We study some examples of its behaviour in Section~\ref{entgt-dyn}.

For Gaussian states, the entropy is not affected by symplectic transformation. If the symplectic eigenvalues are $\lb_k = 2 N_k +1$, where $N_k$ is the number expectation value of the thermal state $\rho_k$  corresponding to the $k$-th normal mode of $\rho$, then the entropy is:
\begin{equation}
S(\rho) = \sum_{k=1}^n (N_k +1) \ln (N_k +1) - N_k \ln N_k\, .
\end{equation}
We note that the entropy is obtained in terms of the symplectic eigenvalues of $\Gm$ without partial transposition. We examine the entropy in Section~\ref{entropy}.

\section{Dynamical and steady state properties of entanglement\label{entgt-dyn}} 

\begin{figure}
\begin{center}
\includegraphics[width= .8\linewidth]{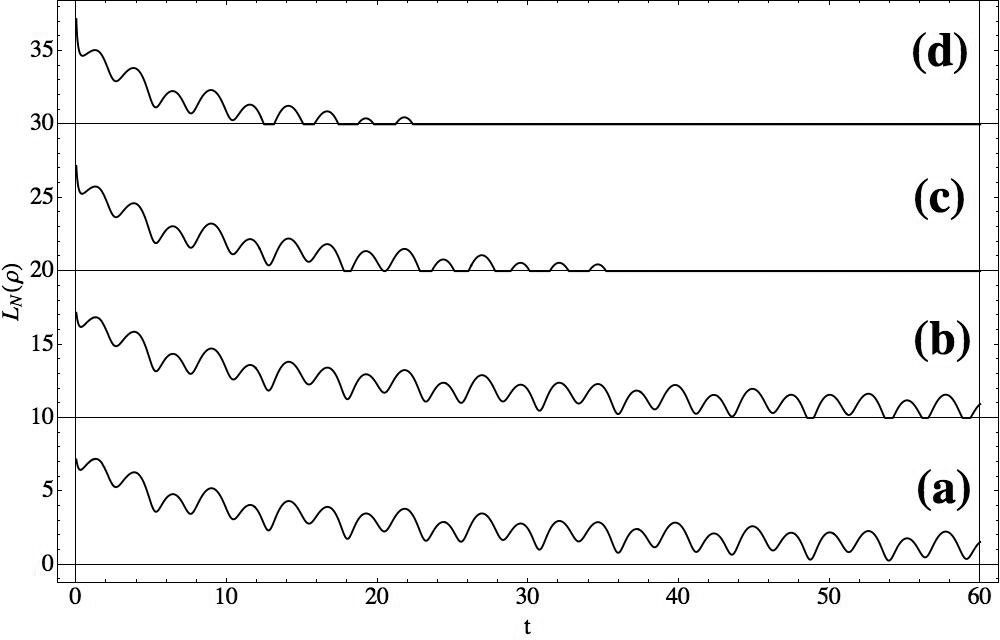}
\caption{Logarithmic negativity versus time for different temperatures of the bath. The parameters of the model are chosen to be $m=2$, $\omega=1$, $\kappa=-1$, $\gamma_1=\gamma_2=0.01$, $s=1$ and $d=6$; (a) shows $L_N(\rho)(t)$ for $T_1=1$ and $T_2=1/4$; (b) shows $L_N(\rho)(t)+10$ for $T_1=1$ and $T_2=1$; (c) shows $L_N(\rho)(t)+20$ for $T_1=1$ and $T_2=4$; (d) shows $L_N(\rho)(t)+30$ for $T_1=4$ and $T_2=4$.}
\end{center}
\end{figure}

In this Section we analyse the dynamical and steady state properties of entanglement between oscillators. For convenience we choose a system of units were $\hbar=k_B=1$.
The dynamics of the entanglement between oscillators for different temperatures of the bath is analysed in Fig. 1. One can clearly see in Figs. 1c and 1d that for higher temperatures of the baths the entanglement vanishes. 
In the low temperature case depicted in Figs. 1a and 1b entanglement persists for longer times. All curves in  Fig. 1 show that there is an interplay between dissipative effects and entanglement created by the harmonic interaction between oscillators. The fact that entanglement disappears at higher temperatures is intuitively understandable. For the free, undriven system at higher temperatures of the bath, average thermal fluctuations are higher than any other characteristic of the system. For the lower temperature thermal fluctuations are comparable with the energy of interaction between oscillators. In this case oscillators will remain entangled in the non-equilibrium steady state. The absence of thermal entanglement in the high temperature case can also be easily understood from the analysis of the steady state. One can find analytically that for the thermal equilibrium case $\gamma_1=\gamma_2$ and $T_1=T_2=T$ the symplectic eigenvalues of the steady state are $\lb_{1,3}^T = \frac{2 (T/\omega)}{\sqrt{1\pm \alpha}}$, where $\alpha$ is a dimensionless parameter characterising the strength of interaction between oscillators, i.e., $\alpha=\frac{\kappa}{m \omega^2}$. Non-zero entanglement corresponds to the case where one of the symplectic eigenvalues is smaller than $1$, $|\lb_{1,3}^T|<1$. However, taking into account physical boundaries on $\alpha$ ($|\alpha|<1$) and the high-temperature limit ($T/\omega\gg 1$) it is clear that in the high-temperature case symplectic eigenvalues are always large ($|\lb_{1,3}^T|\gg1$) and for the long time limit the system is separable.

\begin{figure}
\begin{center}
\includegraphics[width= .8\linewidth]{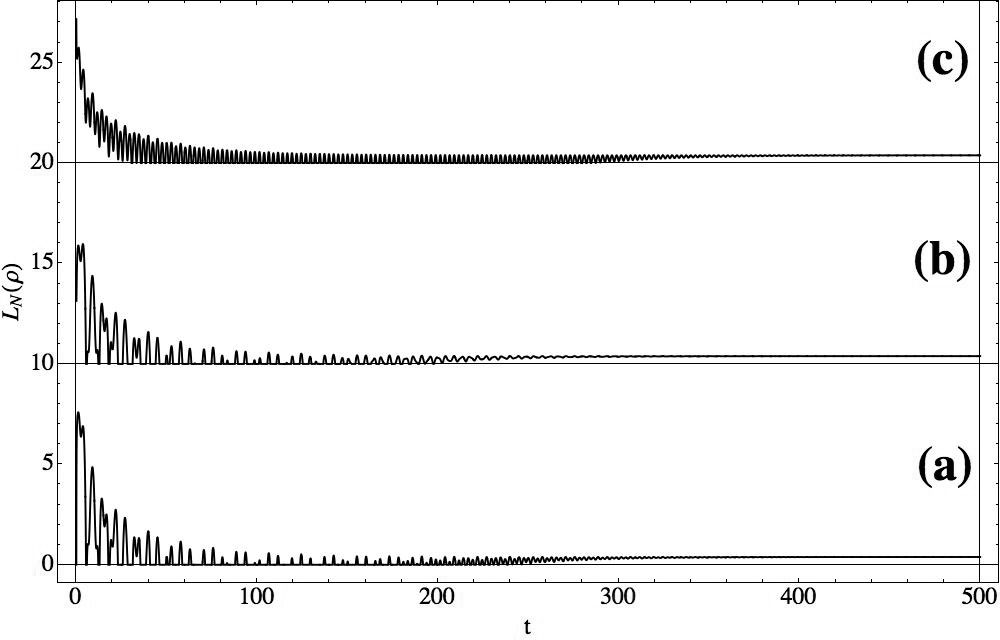}
\caption{\label{Entgt_vary_temp} Logarithmic negativity versus time for different initial conditions of bi-partite system. The parameters of the model are chosen to be $m=2$, $\omega=1$, $\kappa=-1$, $\gamma_1=\gamma_2=0.05$, $T_1=1/3$ and $T_2=1/4$; (a) shows $L_N(\rho)(t)$ for $s=6$ and $d=3$ (separable state); (b) shows $L_N(\rho)(t)+10$ for $s=6$ and $d=1$; (c) shows $L_N(\rho)(t)+20$ for $s=1$ and $d=6$.}
\end{center}
\end{figure}

This may seem to contradict results from Ref. \cite{Anders:2008}, where the entanglement of a thermal state in harmonic oscillator lattices was analysed: a critical temperature was derived, which in the high temperature limit becomes $T_c = \frac{\hb \kappa}{2 k_B}$. Using the parameters from Fig. 1 yields $T_c = 0.5$. It is clear that this value of the critical temperature is much smaller than temperatures applicable for the high temperature limit considered in the present article (the condition $T/\omega\gg 1$ is not satisfied).
Analysing the steady state obtained here, we notice that even in the thermal equilibrium case ($T_1=T_2=T$) the steady state density matrix does not take the form of the canonical ensemble. However, this situation is not surprising and similar situations are studied for spin chains coupled at the ends to heat reservoirs \cite{TP:2010}. If one would like to obtain a canonical distribution in the steady state, there are two ways. The first option is to consider a system bath coupling to a common bath as in Ref. \cite{Chou:2008}. The second option is to consider a position coupling which is transformed to a normal mode coupling using the rotating wave approximation. In this case the density matrix in thermal equilibrium will take a canonical form. A similar behaviour was observed in a spin system \cite{Ilya:2008}.   

Fig. 2 shows the long time dependance of the logarithmic negativity for different initial conditions of the oscillators. In this case we consider a relatively low temperature case so that thermal fluctuations do not cancel the quantum correlations between the oscillators. For times smaller than $10/\gamma$ we can see competition between quantum correlations created by the unitary interaction between oscillators and dissipative effects. For times between $10/\gamma$ and $15/\gamma$ thermal entanglement slowly grows. For all initial conditions at larger times the system reaches a unique non-equilibrium entangled steady state. 

\begin{figure}
\begin{center}
 \includegraphics[width= .9\linewidth]{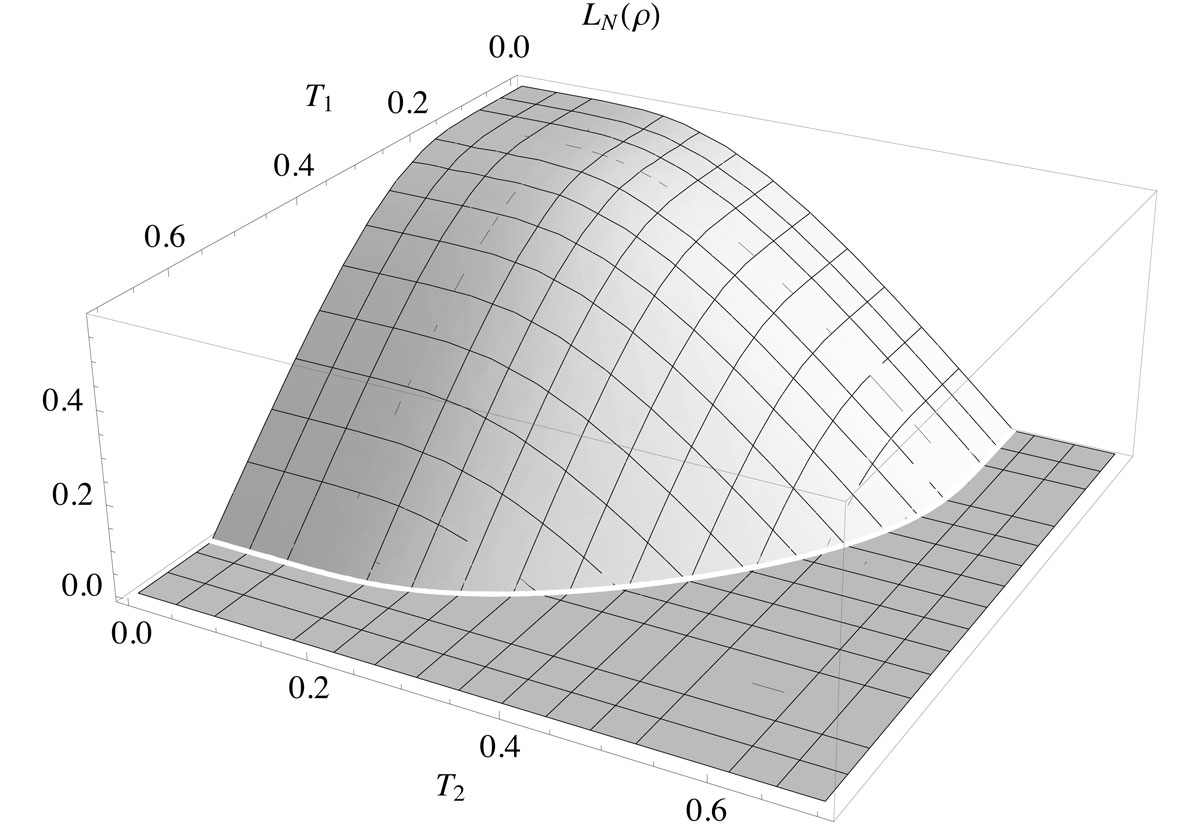}
\caption{Steady state logarithmic negativity as function of the temperatures of the baths. The parameters of the model are chosen to be $m=2$, $\omega=1$, $\kappa=-1$, $\gamma_1=\gamma_2=0.0001$,}
\end{center}
\end{figure}

In Fig. 3 we analyse the non-equilibrium steady state entanglement. It is easy to see that the lower the temperature of the baths, the higher the thermal entanglement. This behaviour is different from a spin system in a similar configuration \cite{Ilya:2008}. For a spin system in a similar configuration entanglement reaches a maximum for a bath at non-zero temperature. Fig. 3. also shows that there is a well-defined border in the temperature plane above which the system is separable. Using the explicit expression for a steady state in the weak coupling limit (Appendix B) and assuming for simplicity that the friction coefficients for both baths are the same, $\gamma_1=\gamma_2=\gamma$ we obtain explicit expressions for the symplectic eigenvalues of the steady state. The explicit expression for $\lambda_{1,3}^T$ are quite large and in order to analyse thermal entanglement we expand the symplectic eigenvalues using a parameter $\gamma$, i.e., $\lambda_{1,3}^T(\gamma)\approx\lambda_{1,3}^T(0)+\gamma\frac{d\lambda_{1,3}^T(\gamma)}{d\gamma}\Big\lvert_{\gamma=0}+O(\gamma^2)$. However, for the situation considered here the term linear in $\gamma$ is $0$, so $\lambda_{1,3}^T(\gamma)\approx\lambda_{1,3}^T(0)+O(\gamma^2)$. Using the weak coupling limit approximation the explicit expression for the symplectic eigenvalues takes the form,
\begin{equation}
\lambda_{1,3}^T\approx\frac{\coth{(\frac{\omega}{2T_1})}+\coth{(\frac{\omega}{2T_2})}}{2\sqrt{1\pm\alpha}}.
\end{equation}
If one considers the thermal equilibrium case and applies the high-temperature limit, one immediately recovers the symplectic eigenvalues obtained earlier for the high-temperature case. One can simplify the explicit expression for the logarithmic negativity using the fact that for all positive $x$ the function $\coth(x)$ will be greater or equal to $1$ ($\forall x>0, \coth(x)\ge 1$). This implies that out of two distinctive eigenvalues $\lambda_{1,3}^T$ at least one should be always bigger than $1$ and does not contribute to the value of the logarithmic negativity. The symplectic eigenvalues which can be smaller than $1$ can be written as,
\begin{equation}
\lambda^T=\frac{\coth{(\frac{\omega}{2T_1})}+\coth{(\frac{\omega}{2T_2})}}{2\sqrt{1+|\alpha|}},
\end{equation}
and the explicit expression for the logarithmic negativity takes the following form,
\begin{equation}\label{negSS}
\mathcal{L_{\mathcal{N}}}(\rho) = -2 \log_2\left(\min(1, \frac{\coth{(\frac{\omega}{2T_1})}+\coth{(\frac{\omega}{2T_2})}}{2\sqrt{1+|\alpha|}})\right).
\end{equation}
The dependence of the steady state logarithmic negativity on the dimensionless strength of the interaction between the oscillators $\alpha=\kappa/m\omega^2$ is presented in Fig. 4. In the case of low temperatures of the baths (Fig. 4c) a very weak interaction between oscillators is enough to create steady state entanglement. While for higher temperatures of the baths (Fig. 4a and 4b) the strength of the oscillator interaction should be higher to create a non-separable steady state. In all cases we can see that the maximum of the thermal entanglement is achieved for stronger interaction $\alpha$ between the oscillators and lower temperatures of the baths.
It follows from Eq. (\ref{negSS}) that the critical temperatures of the baths satisfy
\begin{equation}
\coth{(\frac{\omega}{2T_1^c})}+\coth{(\frac{\omega}{2T_2^c})}=2\sqrt{1+|\alpha|}.
\end{equation}
For the thermal equilibrium case ($T_1=T_2=T$) the critical temperature can be easy calculated,
\begin{equation}
T^c=\frac{\omega}{2\coth^{-1}{\sqrt{1+|\alpha|}}}.
\end{equation}
In Fig. 5 the steady-state logarithmic negativity for the thermal equilibrium case as a function of the temperature of the bath and oscillator coupling strength $\alpha$ is presented. It is interesting to mention that for the range of low temperatures $T<0.15$ the thermal fluctuations are so small that for all values of $\alpha$ there is thermal entanglement. For this low temperature case one can estimate the logarithmic negativity. Using Eq. (\ref{negSS}) one gets, $\mathcal{L_{\mathcal{N}}}(\rho) \approx \log_2\left(1+|\alpha|\right)$.
From Fig. 5 one sees that there is a critical temperature $T^c=1/2\coth^{-1}(\sqrt{2})\approx 0.57$.
 
\begin{figure}
\begin{center}
\includegraphics[scale=0.35]{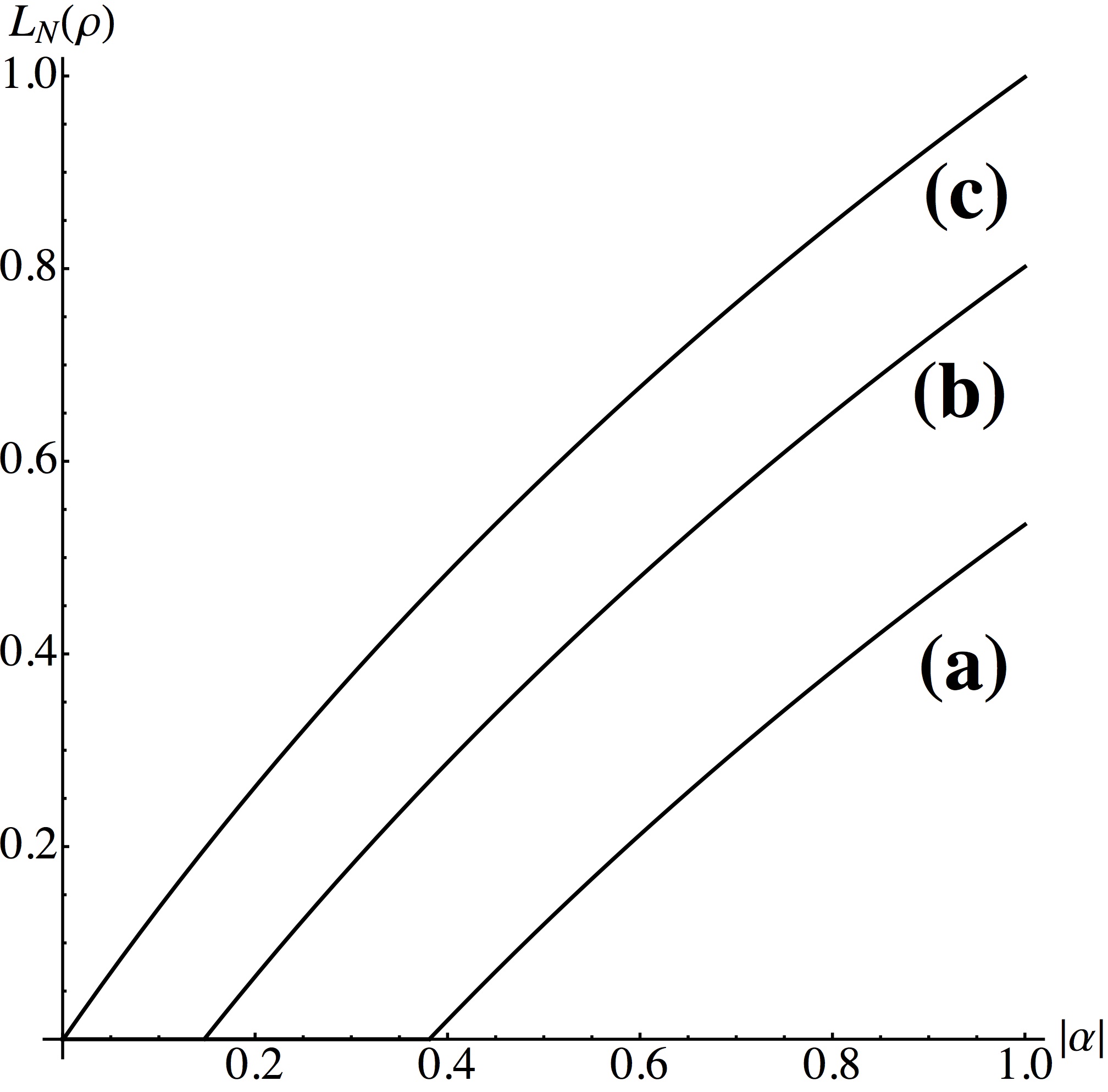}
\caption{\label{Entgt_vary_gam} Steady state logarithmic negativity as function of dimensionless parameter $\alpha$. The parameter of the model is chosen to be $\omega=1$; (a) shows $L_N(\rho)$ for $T_1=1/2$ and $T_2=1/4$; (b) shows $L_N(\rho)$ for $T_1=1/3$ and $T_2=1/4$; (c) shows $L_N(\rho)$ for $T_1=1/10$ and $T_2=1/8$.}
\end{center}
\end{figure}

\begin{figure}
\begin{center}
 \includegraphics[scale=0.35]{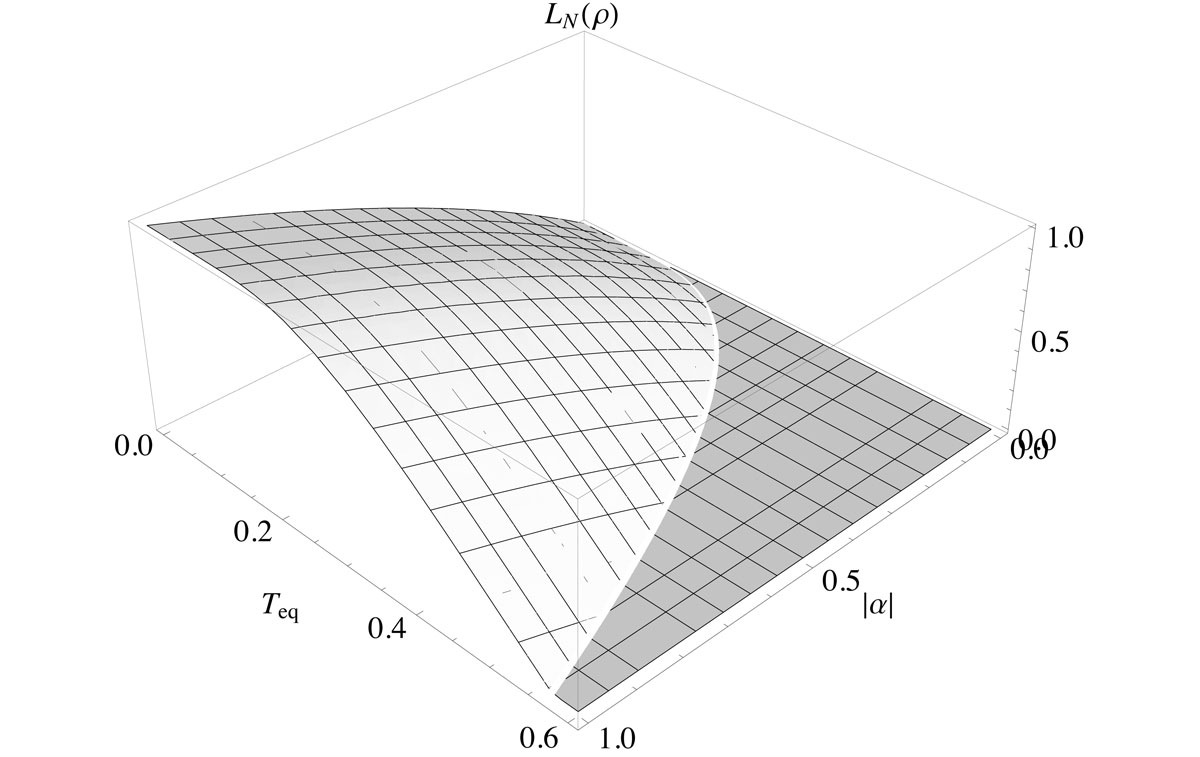}
\caption{\label{Entgt_vary_gam} Steady state logarithmic negativity as function of the equilibrium temperature of the bath and parameter $\alpha$. The parameter of the model is chosen to be $\omega=1$.}
\end{center}
\end{figure}

\section{Entropy dynamics \label{entropy}}
In order to demonstrate the equilibration in the system for the high-temperature case we analyse the dynamics of the entropy.  
\begin{figure}
\begin{center}
\includegraphics[scale=0.45]{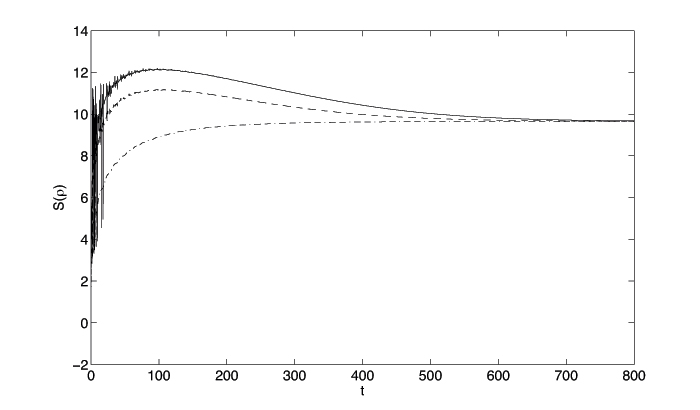}
\caption{\label{Entp_vary_s} Entropy versus time. The bath parameters are chosen to be $T_2=4$, $T_1=2$, $m=1$, $\gm_1=0.009$, $\gm_2=0.01$, $\og_0=1.3$, $\kp= - 1.6$. The system parameters are chosen to be  $d=6$, $s = 10$ (full), $s=6$ (dashed), $s=1$ (dash-dotted).}
\end{center}
\end{figure}

In Fig.~\ref{Entp_vary_s} we show the influence of varying $s$, the distance between the particles, keeping the position of the centre of mass $d$ constant. It is obvious that all curves converge to the same asymptotic value. This indicates that the system reaches the same steady state for all initial conditions. Although not plotted, one observes the same results when keeping the distance $s$ constant and varying $d$. As the distance $s$ increases, one also notices a bump in the entropy. One may, however, note that the time at which the system reaches its non-equilibrium steady state depends on the initial conditions.

\begin{figure}
\begin{center}
\includegraphics[scale=0.45]{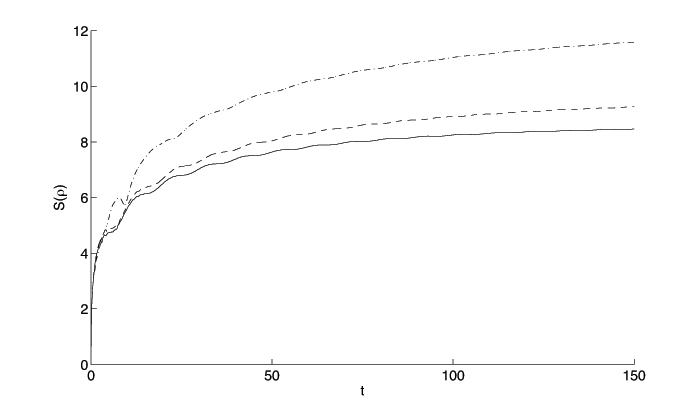}
\caption{ \label{Entp_vary_og}Entropy versus time. The system parameters are chosen to be $d=6$, $s=1$. The bath parameters are chosen to be $T_2=4$, $T_1=2$, $m=1$, $\gm_1=0.009$, $\gm_2=0.01$, $\og_0=2$, $\kp= -3.9$ (full), $\og_0=0.5$, $\kp= -0.2$ (dash-dotted), $\og_0=1.3$, $\kp= -1.6$ (dashed).}
\end{center}
\end{figure}

Fig.~\ref{Entp_vary_og} allows us to examine the approach to the non-equilibrium steady state when one varies the frequency of the oscillators and the strength of the interaction between them. One may observe that when $\og_0$ and $\kp$ are either quite small (full curve, over-damped) or quite large (dash-dotted curve, very under-damped), the system reaches its non-equilibrium steady state quite smoothly.  The oscillations visible on the dashed curve suggest that the system may be more sensitive to such strengths for $\og_0$ and $\kp$. 

\begin{figure}
\begin{center}
\includegraphics[scale=0.45]{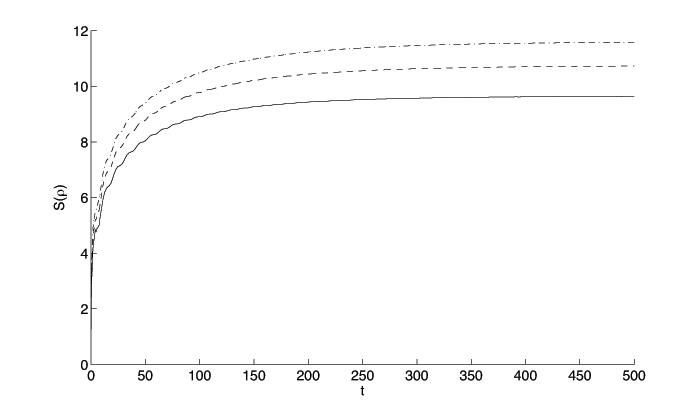}
\caption{\label{Entp_vary_temp} Entropy versus time. The system parameters are chosen to be $d=6$, $s=1$. The bath parameters are chosen to be $T_2=4$, $m=1$, $\gm_1=0.009$, $\gm_2=0.01$, $\og_0=1.3$, $\kp= - 1.6$, $T_1=2$ (full), $T_1=4$ (dashed), $T_1=6$ (dash-dotted).}
\end{center}
\end{figure}

Fig.~\ref{Entp_vary_temp} illustrates how varying the bath's parameters affects the approach to the non-equilibrium steady state. In Figure~\ref{Entp_vary_temp}, one may easily notice that as one would expect, when the temperature of one bath is increased, the system approaches  its non-equilibrium steady state at a higher value of entropy. However, they do so at around the same time. The effect of changing of the friction coefficient $\gamma_i$ will result only in rescaling equilibration time in the system. The smaller the coupling with the bath, the longer it takes for the system to reach its non-equilibrium steady state. Indeed, increasing the coupling or the temperature increases the disturbance that the system suffers from the reservoirs and thus accelerates its loss of energy.

\section{\label{conclusion}Concluding remarks}
A system composed of two entangled particles, each coupled to an independent reservoir, was studied. This set-up enabled us to study the effects of the dissipation induced by the coupling to the reservoirs. It was observed that increasing temperatures dampens the entanglement within the system. Similarly, it was found that a stronger coupling to the reservoir tends to destroy the entanglement. The system was studied in two limit cases, quantum Brownian motion limit ($T/\omega\gg1$) and weak coupling limit ($\gamma\ll \omega,m,\kappa$). It was found that there is no thermal entanglement in the high temperature case. In the weak coupling limit the properties of thermal entanglement and approach to non-equilibrium steady-state were studied. Boundaries for thermal entanglement and critical temperatures are found. The dependence of the thermal entanglement on the coupling strength between the oscillators and temperatures of the baths was studied. In the high temperature case we used entropy as the witness of equilibration of the system. The approach to the non-equilibrium steady state was clearly noticed. It was also observed that the system reaches its non-equilibrium steady state all the more quickly, the higher the temperatures of the bath and the coupling of the system to them, as result of the dissipation induced by the two reservoirs. 

\section*{Acknowledgments} 
This work is based on research supported by the South African Research Chair Initiative of the Department of Science and Technology and National Research Foundation.

\appendix

\section{Analytical solution of the master equation  \label{App_solution}}
The main steps to solving the master equation will be described here. In (\ref{master_equation}), the change of variables $x = u + \hb z$ and $y = u - \hb z$ is performed. Then the Fourier transform is applied
\be {\tilde P}(\tbf{q}, \tbf{z}, t) = \int du_1 \, du_2 \, P(\tbf{u}, \tbf{z}, t) e^{- \im q_1 u_1 - \im q_2 u_2}. \non \ee
This yields
\begin{align} \label{eqtosolve}
\lefteqn{ \frac{\partial {\tilde P}}{\partial t}(\tbf{q}, \tbf{z}, t)} \non 
\\ &= \Biggl\{ - \frac{1}{2 m} \left( q_1 \frac{\partial}{\partial z_1} + q_2 \frac{\partial}{\partial z_2} \right) - \frac{\gm_1}{m} \, z_1 \, \frac{\partial}{\partial z_1} - \frac{\gm_2}{m} \, z_2 \, \frac{\partial}{\partial z_2} \non 
\\ & + 2 (m \og_0 ^2 z_1 + \kp z_2) \frac{\partial}{\partial q_1} + 2 (m \og_0 ^2 z_2 + \kp z_1) \frac{\partial}{\partial q_2}  \non
\\ &- 4 \gm_1 k T_1 \, z_1 ^2 - 4 \gm_2 k T_2 \, z_2 ^2  \Biggr\} {\tilde P}(\tbf{q}, \tbf{z}, t),  
\end{align}
which can be solved using the method of characteristics. The characteristic equation yields
\be \frac{\partial \tbf{v}}{\partial t} = \frac{M}{2m} \tbf{v}, \non \ee
with  $\tbf{v}^T = (z_1, z_2, q_1, q_2)$ and 
\be M= \left( \begin{array}{cccc} 2\gm_1 & 0 & 1 & 0 \\ 0 & 2\gm_2 & 0 & 1 \\ - 4 m^2 \og_0 ^2 & - 4 m \kp & 0 & 0 \\ - 4 m \kp & - 4 m^2 \og_0 ^2 & 0 & 0 \end{array} \right), \non \ee
so that on a characteristic, 
\be \label{charac} \frac{d {\tilde P}}{d t} = - 4 k( \gm_1 T_1 z_1 ^2 + \gm_2 T_2 z_2 ^2) {\tilde P}. \non \ee
The eigenvalues and eigenvectors of M can be computed to be 
\begin{align} \bds{\lb}^T &= \left(\lb_1, \lb_2, \lb_3, \lb_4 \right)\end{align}
and
\be Q= \left( \begin{array}{cccc} a_1 & a_2 & a_3 & a_4
\\ b_1 & b_2 & b_3 & b_4
\\ c_1 & c_2 & c_3 & c_4
\\ f_1 & f_2 & f_3 & f_4 \end{array}
 \right) \quad \text{and} \quad  Q^{-1} = \left( \begin{array}{cccc} \tilde{a_1} & \tilde{a_2} & \tilde{a_3} & \tilde{a_4} 
\\ \tilde{b_1} & \tilde{b_2} & \tilde{b_3} & \tilde{b_4} 
\\ \tilde{c_1} & \tilde{c_2} & \tilde{c_3} & \tilde{c_4} 
\\ \tilde{f_1} & \tilde{f_2} & \tilde{f_3} & \tilde{f_4} 
\end{array}\right) \non
 \ee
with
\begin{align}
a_i =& \frac{c_i}{\lb_i - 2 \gm_1} ,\non
\\ b_i =& \frac{f_i}{\lb_i - 2 \gm_2},
\\ c_i =& - \frac{\lb_i - 2 \gm_1}{\lb_i - 2 \gm_2} \frac{\lb_i(\lb_i - 2 \gm_2) + 4 m ^2 \og_0 ^2 + 4 m \kp}{\lb_i(\lb_i - 2 \gm_1) + 4 m ^2 \og_0 ^2 + 4 m \kp}, \non
\\ f_i =& f_i .\non
\end{align}
We choose $f_i =1$ for simplicity. Using that $Q^{-1} M Q = D$ where D is the diagonal matrix, the characteristic equation is solved, then the master equation.

Using these results, one can rewrite the differential equation as
\be 2 m \frac{\partial \tbf{v}}{\partial t} = Q D Q^{-1} \tbf{v}\, . \non\ee
This is easily solved  and yields
\be \tbf{v}(t) = Q e^{D t / 2m} Q^{-1} \tbf{v}_0. \non \ee
We get $\tbf{v}(t)$
{\alld
\be
\left( \begin{array}{c} \zt_1^z {z_1}_0 + \zt_2^z {z_2}_0 + \zt_1^q {q_1}_0 + \zt_2^q {q_2}_0
\\  \xi_1^z {z_1}_0 + \xi_2^z {z_2}_0 + \xi_1^q {q_1}_0 + \xi_2^q {q_2}_0 
\\  \tau_1^z {z_1}_0 + \tau_2^z {z_2}_0 + \tau_1^q {q_1}_0 + \tau_2^q {q_2}_0 
\\  \vta_1^z {z_1}_0 + \vta_2^z {z_2}_0 + \vta_1^q {q_1}_0 + \vta_2^q {q_2}_0 \end{array} \right) \non
\ee}
with
{\alld
\begin{align}
\zt_{1,2} ^z = a_1 \tilde{a_{1,2}} e^{\lb_1 t /2m} &+ a_2 \tilde{b_{1,2}} e^{\lb_2 t / 2m} + a_3 \tilde{c_{1,2}} e^{\lb_3 t / 2m} \non
\\ &+ a_4 \tilde{f_{1,2}} e^{\lb_4 t /2m}, \non
\\ \zt_{1,2} ^q = a_1 \tilde{a_{3,4}} e^{\lb_1 t /2m} &+ a_2 \tilde{b_{3,4}} e^{\lb_2 t / 2m} + a_3 \tilde{c_{3,4}} e^{\lb_3 t / 2m} \non
\\ &+ a_4 \tilde{f_{3,4}} e^{\lb_4 t /2m} ,\non
\\ \xi_{1,2} ^z = b_1 \tilde{a_{1,2}} e^{\lb_1 t /2m} &+ b_2 \tilde{b_{1,2}} e^{\lb_2 t / 2m} + b_3 \tilde{c_{1,2}} e^{\lb_3 t / 2m} \non
\\ &+ b_4 \tilde{f_{1,2}} e^{\lb_4 t /2m}, \non
\\ \xi_{1,2} ^q = b_1 \tilde{a_{3,4}} e^{\lb_1 t /2m} &+ b_2 \tilde{b_{3,4}} e^{\lb_2 t / 2m} + b_3 \tilde{c_{3,4}} e^{\lb_3 t / 2m} \non
\\ &+ b_4 \tilde{f_{3,4}} e^{\lb_4 t /2m}, \non
\\ \tau_{1,2} ^z = c_1 \tilde{a_{1,2}} e^{\lb_1 t /2m} &+ c_2 \tilde{b_{1,2} } e^{\lb_2 t / 2m} + c_3 \tilde{c_{1,2} } e^{\lb_3 t / 2m} \non
\\ &+ c_4 \tilde{f_{1,2} } e^{\lb_4 t /2m}, \non 
\\ \tau_{1,2}  ^q = c_1 \tilde{a_{3,4}} e^{\lb_1 t /2m} &+ c_2 \tilde{b_{3,4}} e^{\lb_2 t / 2m} + c_3 \tilde{c_{3,4}} e^{\lb_3 t / 2m} \non
\\ &+ c_4 \tilde{f_{3,4}} e^{\lb_4 t /2m} ,\non
\\ \vta_{1,2}  ^z = f_1 \tilde{a_{1,2} } e^{\lb_1 t /2m} &+ f_2 \tilde{b_{1,2} } e^{\lb_2 t / 2m} + f_3 \tilde{c_{1,2} } e^{\lb_3 t / 2m} \non
\\ &+ f_4 \tilde{f_{1,2} } e^{\lb_4 t /2m}, \non
\\ \vta_{1,2}  ^q = f_1 \tilde{a_{3,4}} e^{\lb_1 t /2m} &+ f_2 \tilde{b_{3,4}} e^{\lb_2 t / 2m} + f_3 \tilde{c_{3,4}} e^{\lb_3 t / 2m} \non
\\ &+ f_4 \tilde{f_{3,4}} e^{\lb_4 t /2m}. \non
\end{align}}
A short note on the notation must now be done. In order to avoid multiplying the cumbersome equations, those which are similar have been condensed and indices (or superscripts as will be found in the following expressions) are used. An example is given by way of explanation. In
\begin{align}
\zt_{1,2} ^z = a_1 \tilde{a_{1,2}} e^{\lb_1 t /2m} &+ a_2 \tilde{b_{1,2}} e^{\lb_2 t / 2m} + a_3 \tilde{c_{1,2}} e^{\lb_3 t / 2m} \non
\\ &+ a_4 \tilde{f_{1,2}} e^{\lb_4 t /2m} \non \, ,\end{align}
the index $1$ in $\zt_{1,2} ^z$ means that to obtain its expression, one must consider the index $1$ in the terms on the right-hand-side as such
\begin{align}
 a_1 \tilde{a_{1}} e^{\lb_1 t /2m} &+ a_2 \tilde{b_{1}} e^{\lb_2 t / 2m} + a_3 \tilde{c_{1}} e^{\lb_3 t / 2m} \non
\\ &+ a_4 \tilde{f_{1}} e^{\lb_4 t /2m} \non \, ;\end{align}
on the other hand, should we need $\zt_2 ^z$, the index $2$ must be considered. 

 We can then insert $z_1 ^2(t)$ and $z_2 ^2(t)$ into (\ref{charac}). After integration, one can write (dropping the t-dependence to keep it readable)
%\begin{widetext}
\begin{align}
\tilde{P}(\tbf{q},\tbf{z},t) = & \tilde{P}_0 \exp \left( - 4 k \left(\chi_1 ^z {z_1}_0 ^2  + \chi_2 ^z {z_2}_0 ^2  + \chi_1 ^q {q_1}_0 ^2 + \chi_2 ^q {q_2}_0 ^2 + \ta ^z {z_1}_0 {z_2}_0 + \ta ^q {q_1}_0 {q_2}_0 \right) \right) \non
\\ & \times \exp \left( - 4 k \left( \Lb_{11} {z_1}_0 {q_1}_0 + \Lb_{12} {z_1}_0 {q_2}_0 + \Lb_{21} {z_2}_0 {q_1}_0 + \Lb_{22} {z_2}_0 {q_2}_0 \right) \right)
\end{align}
with
{\alld
\begin{align}
\chi_1 ^z, \, \chi_2 ^z, \, \chi_1 ^q, \, \chi_2 ^q &= \frac{m\, \tilde{a}_{1,2,3,4} ^2}{\lb_1}(e^{\lb_1 t /m} -1)(\gm_1 T_1 a_1 ^2  + \gm_2 T_2 b_1 ^2) \non
\\ & + \frac{m\, \tilde{b}_{1,2,3,4} ^2}{\lb_2}(e^{\lb_2 t /m} -1)(\gm_1 T_1 a_2 ^2 + \gm_2 T_2 b_2 ^2) \non
\\ & + \frac{m\, \tilde{c}_{1,2,3,4} ^2}{\lb_3}(e^{\lb_3 t /m} -1)(\gm_1 T_1 a_3 ^2  + \gm_2 T_2 b_3 ^2) \non
\\ & + \frac{m\, \tilde{f}_{1,2,3,4} ^2}{\lb_4}(e^{\lb_4 t /m} -1)(\gm_1 T_1 a_4 ^2 + \gm_2 T_2 b_4 ^2) \non
\\ & + \frac{4m\, \tilde{a}_{1,2,3,4} \tilde{b}_{1,2,3,4}}{\lb_1 + \lb_2}(e^{(\lb_1 + \lb_2) t /2m} -1)(\gm_1 T_1 a_1 a_2 + \gm_2 T_2 b_1 b_2 )\non
\\ &+  \frac{4m\, \tilde{a}_{1,2,3,4} \tilde{c}_{1,2,3,4}}{\lb_1 + \lb_3}(e^{(\lb_1 + \lb_3) t /2m} -1)(\gm_1 T_1 a_1 a_3 + \gm_2 T_2 b_1 b_3) \non
\\ & + \frac{4m\, \tilde{a}_{1,2,3,4} \tilde{f}_{1,2,3,4}}{\lb_1 + \lb_4}(e^{(\lb_1 + \lb_4) t /2m} -1)(\gm_1 T_1 a_1 a_4 + \gm_2 T_2 b_1 b_4) \non
\\ & + \frac{4m\, \tilde{b}_{1,2,3,4} \tilde{c}_{1,2,3,4}}{\lb_2 + \lb_3}(e^{(\lb_2 + \lb_3) t /2m} -1)(\gm_1 T_1 a_2 a_3 + \gm_2 T_2 b_2 b_3) \non
\\ & + \frac{4m\, \tilde{b}_{1,2,3,4} \tilde{f}_{1,2,3,4}}{\lb_2 + \lb_4}(e^{(\lb_2 + \lb_4) t /2m} -1)(\gm_1 T_1 a_2 a_4 + \gm_2 T_2 b_2 b_4) \non
\\ & + \frac{4m\, \tilde{c}_{1,2,3,4} \tilde{f}_{1,2,3,4}}{\lb_3 + \lb_4}(e^{(\lb_3 + \lb_4) t /2m} -1)(\gm_1 T_1 a_3 a_4 + \gm_2 T_2 b_3 b_4),\non \\ \non
\\ \ta^z, \, \ta^q &= \frac{2m\, \tilde{a}_{1,3} \tilde{a}_{2,4}}{\lb_1}(e^{\lb_1 t /m} -1)(\gm_1 T_1 a_1 ^2  + \gm_2 T_2 b_1 ^2) \non
\\ & + \frac{2m\, \tilde{b}_{1,3} \tilde{b}_{2,4}}{\lb_2}(e^{\lb_2 t /m} -1)(\gm_1 T_1 a_2 ^2 + \gm_2 T_ 2 b_2 ^2) \non
\\ & + \frac{2m\, \tilde{c}_{1,3} \tilde{c}_{2,4}}{\lb_3}(e^{\lb_3 t /m} -1)(\gm_1 T_1 a_3 ^2  + \gm_2 T_2 b_3 ^2) \non
\\ & + \frac{2m\, \tilde{f}_{1,3} \tilde{f}_{2,4}}{\lb_4}(e^{\lb_4 t /m} -1)(\gm_1 T_1 a_4 ^2 + \gm_2 T_2 b_4 ^2) \non
\\ & + \frac{4m\, ( \tilde{a}_{1,3} \tilde{b}_{2,4} + \tilde{b}_{1,3} \tilde{a}_{2,4} ) }{\lb_1 + \lb_2}(e^{(\lb_1 + \lb_2) t /2m} -1)(\gm_1 T_1 a_1 a_2 + \gm_2 T_2 b_1 b_2)\non
\\ &+  \frac{4m\, ( \tilde{a}_{1,3} \tilde{c}_{2,4} + \tilde{c}_{1,3} \tilde{a}_{2,4} )}{\lb_1 + \lb_3}(e^{(\lb_1 + \lb_3) t /2m} -1)(\gm_1 T_1 a_1 a_3 + \gm_2 T_2 b_1 b_3) \non
\\ & + \frac{4m\, ( \tilde{a}_{1,3} \tilde{f}_{2,4} + \tilde{f}_{1,3} \tilde{a}_{2,4} )}{\lb_1 + \lb_4}(e^{(\lb_1 + \lb_4) t /2m} -1)(\gm_1 T_1 a_1 a_4 + \gm_2 T_2 b_1 b_4) \non
\\ & + \frac{4m\, ( \tilde{b}_{1,3} \tilde{c}_{2,4} + \tilde{c}_{1,3} \tilde{b}_{2,4} )}{\lb_2 + \lb_3}(e^{(\lb_2 + \lb_3) t /2m} -1)(\gm_1 T_1 a_2 a_3 + \gm_2 T_2 b_2 b_3) \non
\\ & + \frac{4m\, ( \tilde{b}_{1,3} \tilde{f}_{2,4} + \tilde{f}_{1,3} \tilde{b}_{2,4} )}{\lb_2 + \lb_4}(e^{(\lb_2 + \lb_4) t /2m} -1)(\gm_1 T_1 a_2 a_4 + \gm_2 T_2 b_2 b_4) \non
\\ & + \frac{4m\, ( \tilde{c}_{1,3} \tilde{f}_{2,4} + \tilde{f}_{1,3} \tilde{c}_{2,4} )}{\lb_3 + \lb_4}(e^{(\lb_3 + \lb_4) t /2m} -1)(\gm_1 T_1 a_3 a_4 + \gm_2 T_2 b_3 b_4),\non \\ \non
\\  \Lb_{11}, \, \Lb_{12} &= \frac{2m\, \tilde{a}_1 \tilde{a}_{3,4}}{\lb_1}(e^{\lb_1 t /m} -1)(\gm_1 T_1 a_1 ^2  + \gm_2 T_2 b_1 ^2) \non
\\ & + \frac{2m\, \tilde{b}_1 \tilde{b}_{3,4}}{\lb_2}(e^{\lb_2 t /m} -1)(\gm_1 T_1 a_2 ^2 + \gm_2 T_2 b_2 ^2) \non
\\ & + \frac{2m\, \tilde{c}_1 \tilde{c}_{3,4}}{\lb_3}(e^{\lb_3 t /m} -1)(\gm_1 T_1 a_3 ^2  + \gm_2 T_2 b_3 ^2) \non
\\ & + \frac{2m\, \tilde{f}_1 \tilde{f}_{3,4}}{\lb_4}(e^{\lb_4 t /m} -1)(\gm_1 T_1 a_4 ^2 + \gm_2 T_2 b_4 ^2) \non
\\ & + \frac{4m\, (\tilde{a}_1 \tilde{b}_{3,4} + \tilde{b}_1 \tilde{a}_{3,4})}{\lb_1 + \lb_2}(e^{(\lb_1 + \lb_2) t /2m} -1)(\gm_1 T_1 a_1 a_2 + \gm_2 T_2 b_1 b_2 )\non
\\ &+  \frac{4m\, (\tilde{a}_1 \tilde{c}_{3,4} + \tilde{c}_1 \tilde{a}_{3,4})}{\lb_1 + \lb_3}(e^{(\lb_1 + \lb_3) t /2m} -1)(\gm_1 T_1 a_1 a_3 + \gm_2 T_2 b_1 b_3) \non
\\ & + \frac{4m\, (\tilde{a}_1 \tilde{f}_{3,4} + \tilde{f}_1 \tilde{a}_{3,4} )}{\lb_1 + \lb_4}(e^{(\lb_1 + \lb_4) t /2m} -1)(\gm_1 T_1 a_1 a_4 + \gm_2 T_2 b_1 b_4) \non
\\ & + \frac{4m\, (\tilde{b}_1 \tilde{c}_{3,4} + \tilde{c}_1 \tilde{b}_{3,4} )}{\lb_2 + \lb_3}(e^{(\lb_2 + \lb_3) t /2m} -1)(\gm_1 T_1 a_2 a_3 + \gm_2 T_2 b_2 b_3) \non
\\ & + \frac{4m\, (\tilde{b}_1 \tilde{f}_{3,4} + \tilde{f}_1 \tilde{b}_{3,4} )}{\lb_2 + \lb_4}(e^{(\lb_2 + \lb_4) t /2m} -1)(\gm_1 T_1 a_2 a_4 + \gm_2 T_2 b_2 b_4) \non
\\ & + \frac{4m\, (\tilde{c}_1 \tilde{f}_{3,4} + \tilde{f}_1 \tilde{c}_{3,4} )}{\lb_3 + \lb_4}(e^{(\lb_3 + \lb_4) t /2m} -1)(\gm_1 T_1 a_3 a_4 + \gm_2 T_2 b_3 b_4),\non \\ \non
\\ \Lb_{21}, \, \Lb_{22} &= \frac{2m\, \tilde{a}_2 \tilde{a}_{3,4}}{\lb_1}(e^{\lb_1 t /m} -1)(\gm_1 T_1 a_1 ^2  + \gm_2 T_2 b_1 ^2) \non
\\ & + \frac{2m\, \tilde{b}_2 \tilde{b}_{3,4}}{\lb_2}(e^{\lb_2 t /m} -1)(\gm_1 T_1 a_2 ^2 + \gm_2 T_2 b_2 ^2) \non
\\ & + \frac{2m\, \tilde{c}_2 \tilde{c}_{3,4}}{\lb_3}(e^{\lb_3 t /m} -1)(\gm_1 T_1 x_3 ^2  + \gm_2 T_2 b_3 ^2) \non
\\ & + \frac{2m\, \tilde{f}_2 \tilde{f}_{3,4}}{\lb_4}(e^{\lb_4 t /m} -1)(\gm_1 T_1 x_4 ^2 + \gm_2 T_2 b_4 ^2) \non
\\ & + \frac{4m\, (\tilde{a}_2 \tilde{b}_{3,4} + \tilde{b}_2 \tilde{a}_{3,4})}{\lb_1 + \lb_2}(e^{(\lb_1 + \lb_2) t /2m} -1)(\gm_1 T_1 a_1 a_2 + \gm_2 T_2 b_1 b_2 )\non
\\ &+  \frac{4m\, (\tilde{a}_2 \tilde{c}_{3,4} + \tilde{c}_2 \tilde{a}_{3,4})}{\lb_1 + \lb_3}(e^{(\lb_1 + \lb_3) t /2m} -1)(\gm_1 T_1 a_1 a_3 + \gm_2 T_2 b_1 b_3) \non
\\ & + \frac{4m\, (\tilde{a}_2 \tilde{f}_{3,4} + \tilde{f}_2 \tilde{a}_{3,4} )}{\lb_1 + \lb_4}(e^{(\lb_1 + \lb_4) t /2m} -1)(\gm_1 T_1 a_1 a_4 + \gm_2 T_2 b_1 b_4) \non
\\ & + \frac{4m\, (\tilde{b}_2 \tilde{c}_{3,4} + \tilde{c}_2 \tilde{b}_{3,4} )}{\lb_2 + \lb_3}(e^{(\lb_2 + \lb_3) t /2m} -1)(\gm_1 T_1 a_2 a_3 + \gm_2 T_2 b_2 b_3) \non
\\ & + \frac{4m\, (\tilde{b}_2 \tilde{f}_{3,4} + \tilde{f}_2 \tilde{b}_{3,4} )}{\lb_2 + \lb_4}(e^{(\lb_2 + \lb_4) t /2m} -1)(\gm_1 T_1 a_2 a_4 + \gm_2 T_2 b_2 b_4) \non
\\ & + \frac{4m\, (\tilde{c}_2 \tilde{f}_{3,4} + \tilde{f}_2 \tilde{c}_{3,4} )}{\lb_3 + \lb_4}(e^{(\lb_3 + \lb_4) t /2m} -1)(\gm_1 T_1 a_3 a_4 + \gm_2 T_2 b_3 b_4).\non \\ \non
\end{align}}
The Fourier transform of the initial state ($\hb = 1$) is written as
\begin{align}
 {\tilde P}(\tbf{q}_0,\tbf{z}_0;0) &= \exp\left[-\ep_{+} {z_1}_0 ^2-\ep_{+}  {z_2}_0 ^2 + 2\ep_{-} {z_1}_0 {z_2}_0 -\tilde{\ep}_{+}{q_2}_0 ^2 -\tilde{\ep}_{+}{q_1}_0 ^2-2 \tilde{\ep}_{-}{q_1}_0 {q_2}_0 \right]\, ,
\end{align} 
so the final solution yields
{\alld
\begin{align}
\tilde{P}(\tbf{q},\tbf{z},t) =& \exp\left(- \mathcal{A}_1 q_1 ^2 - \mathcal{A}_2 q_2 ^2 - \mathcal{B}_1 z_1 ^2 - \mathcal{B}_2 z_2 ^2 - \mathcal{E} q_1 q_2 - \mathcal{D} z_1 z_2\right)\non
\\ & \times \exp \left( - \mathcal{C}_{11} z_1 q_1 - \mathcal{C}_{22} z_2 q_2 - \mathcal{C}_{12} z_1 q_2 - \mathcal{C}_{21} z_2 q_1 \right)
\end{align}}
where
{\alld
\begin{align}
\mathcal{A}_{1,2} =& (\ep_+ + 4 k \chi_1 ^z) {\zt_{1,2} ^{q_-}} ^2 + (\ep_+ + 4 k \chi_2 ^z) {\xi_{1,2} ^{q_-}} ^2 + (\tilde{\ep}_+ + 4 k \chi_1 ^q) {\tau_{1,2} ^{q_-}} ^2 \non
\\ & + (\tilde{\ep}_+ + 4 k \chi_2 ^q) {\vta_{1,2} ^{q_-}} ^2 - (2 \ep_- - 4 k \ta ^z) {\zt_{1,2} ^{q_-}}{\xi_{1,2} ^{q_-}} + (2 \tilde{\ep}_- + 4 k \ta ^q) {\tau_{1,2} ^{q_-}}{\vta_{1,2} ^{q_-}} \non
\\ & +4 k \left( \Lb_{11} \zt_{1,2} ^{q_-} \tau_{1,2} ^{q_-} + \Lb_{12} \zt_{1,2} ^{q_-} \vta_{1,2} ^{q_-} + \Lb_{21} \xi_{1,2} ^{q_-} \tau_{1,2} ^{q_-} + \Lb_{22} \xi_{1,2} ^{q_-} \vta_{1,2} ^{q_-} \right) ,\non \\ \non
\\ \mathcal{B}_{1,2} =& (\ep_+ + 4 k \chi_1 ^z) {\zt_{1,2} ^{z_-}} ^2 + (\ep_+ + 4 k \chi_2 ^z) {\xi_{1,2} ^{z_-}} ^2 + (\tilde{\ep}_+ + 4 k \chi_1 ^q) {\tau_{1,2} ^{z_-}} ^2 \non
\\ & + (\tilde{\ep}_+ + 4 k \chi_2 ^q) {\vta_{1,2} ^{z_-}} ^2 - (2 \ep_- - 4 k \ta ^z) {\zt_{1,2} ^{z_-}}{\xi_{1,2} ^{z_-}} + (2 \tilde{\ep}_- + 4 k \ta ^q) {\tau_{1,2} ^{z_-}}{\vta_{1,2} ^{z_-}} \non
\\ & +4 k \left( \Lb_{11} \zt_{1,2} ^{z_-} \tau_{1,2} ^{z_-} + \Lb_{12} \zt_{1,2} ^{z_-} \vta_{1,2} ^{z_-} + \Lb_{21} \xi_{1,2} ^{z_-} \tau_{1,2} ^{z_-} + \Lb_{22} \xi_{1,2} ^{z_-} \vta_{1,2} ^{z_-} \right) ,\non \\ \non
\\ \mathcal{D}, \mathcal{E}  =&  2 (\ep_+ + 4 k \chi_1 ^z) \zt_1 ^{z,q_-} {\zt_2 ^{z,q_-}} + 2 (\ep_+ + 4 k \chi_2 ^z) {\xi_1 ^{z,q_-}}  \xi_2 ^{z,q_-} + 2  (\tilde{\ep}_+ + 4 k \chi_1 ^q) \tau_1 ^{z,q_-}  {\tau_2 ^{z,q_-}} \non
\\ & + 2 (\tilde{\ep}_+ + 4 k \chi_2 ^q) {\vta_1 ^{z,q_-}} \vta_2 ^{z,q_-} - (2 \ep_- - 4 k \ta ^z) ({\zt_1 ^{z,q_-}}{\xi_2 ^{z,q_-}} + \zt_2 ^{z,q_-} \xi_1 ^{z,q_-}) \non
\\ & + (2 \tilde{\ep}_- + 4 k \ta ^q) ({\tau_1 ^{z,q_-}}{\vta_2 ^{z,q_-}} + \tau_2 ^{z,q_-} \vta_1 ^{z,q_-}) +4 k \Lb_{11} (\zt_1 ^{z,q_-} \tau_2 ^{z,q_-} + \zt_2 ^{z,q_-} \tau_1 ^{z,q_-} ) \non
\\ & + 4 k (\Lb_{12} (\zt_1 ^{z,q_-} \vta_2 ^{z,q_-} + \zt_2 ^{z,q_-} \vta_1 ^{z,q_-}) + \Lb_{21} (\xi_1 ^{z,q_-} \tau_2 ^{z,q_-} + \xi_2 ^{z,q_-} \tau_1 ^{z,q_-})) \non 
\\ & + 4 k (\Lb_{22} (\xi_1 ^{z,q_-} \vta_2 ^{z,q_-} + \xi_2 ^{z,q_-} \vta_1 ^{z,q_-})), \non \\ \non
\\ \mathcal{C}_{11, 12} =&  2  (\ep_+ + 4 k \chi_1 ^z) \zt_1 ^{z_-} {\zt_{1,2} ^{q_-}} + 2 (\ep_+ + 4 k \chi_2 ^z) {\xi_1 ^{z_-}}  \xi_{1,2} ^{q_-} + 2  (\tilde{\ep}_+ + 4 k \chi_1 ^q) \tau_1 ^{z_-}  {\tau_{1,2} ^{q_-}} \non
\\ & + 2 (\tilde{\ep}_+ + 4 k \chi_2 ^q) {\vta_1 ^{z_-}} \vta_{1,2} ^{q_-} - (2 \ep_- - 4 k \ta ^z) ({\zt_1 ^{z_-}}{\xi_{1,2} ^{q_-}} + \zt_{1,2} ^{q_-} \xi_1 ^{z_-}) \non
\\ & + (2 \tilde{\ep}_- + 4 k \ta ^q) ({\tau_1 ^{z_-}}{\vta_{1,2} ^{q_-}} + \tau_{1,2} ^{q_-} \vta_1 ^{z_-}) + 4 k \Lb_{11} (\zt_1 ^{z_-} \tau_{1,2} ^{q_-} + \zt_{1,2} ^{q_-} \tau_1 ^{z_-} ) \non
\\ & + 4 k (\Lb_{12} (\zt_1 ^{z_-} \vta_{1,2} ^{q_-} + \zt_{1,2} ^{q_-} \vta_1 ^{z_-}) + \Lb_{21} (\xi_1 ^{z_-} \tau_{1,2} ^{q_-} + \xi_{1,2} ^{q_-} \tau_1 ^{z_-}) + \Lb_{22} (\xi_1 ^{z_-} \vta_{1,2} ^{q_-} + \xi_{1,2} ^{q_-} \vta_1 ^{z_-})), \non \\ \non
\\ \mathcal{C}_{21,22} =&  2  (\ep_+ + 4 k \chi_1 ^z) \zt_2 ^{z_-} {\zt_{1,2} ^{q_-}} + 2 (\ep_+ + 4 k \chi_2 ^z) {\xi_2 ^{z_-}}  \xi_{1,2} ^{q_-} + 2  (\tilde{\ep}_+ + 4 k \chi_1 ^q) \tau_2 ^{z_-}  {\tau_{1,2} ^{q_-}} \non
\\ & + 2 (\tilde{\ep}_+ + 4 k \chi_2 ^q) {\vta_2 ^{z_-}} \vta_{1,2} ^{q_-} - (2 \ep_- - 4 k \ta ^z) ({\zt_2 ^{z_-}}{\xi_{1,2} ^{q_-}} + \zt_{1,2} ^{q_-} \xi_2 ^{z_-}) \non
\\ & + (2 \tilde{\ep}_- + 4 k \ta ^q) ({\tau_2 ^{z_-}}{\vta_{1,2} ^{q_-}} + \tau_{1,2} ^{q_-} \vta_2^{z_-}) + 4 k \Lb_{11} (\zt_2 ^{z_-} \tau_{1,2} ^{q_-} + \zt_{1,2} ^{q_-} \tau_2 ^{z_-} ) \non
\\ & + 4 k (\Lb_{12} (\zt_2 ^{z_-} \vta_{1,2} ^{q_-} + \zt_{1,2} ^{q_-} \vta_2 ^{z_-}) + \Lb_{21} (\xi_2 ^{z_-} \tau_{1,2} ^{q_-} + \xi_{1,2} ^{q_-} \tau_2 ^{z_-}) + \Lb_{22} (\xi_2 ^{z_-} \vta_{1,2} ^{q_-} + \xi_{1,2} ^{q_-} \vta_2 ^{z_-})) ,\non \\ \non
\end{align}}
where $\ep_{\pm} = \frac{1}{2 s^2} \pm \frac{1}{8 d^2}$, $\tilde{\ep}_{\pm} = \frac{\ep_{\pm}}{4(\ep_{+} ^2-\ep_{-} ^2)}$, $\zt_i ^{z,q_-} = \zt_i ^{z,q} (-t)$ and similarly for $\xi_i ^{z,q_-}$, $\tau_i ^{z,q_-}$, $\vta_i ^{z,q_-}$.

\section{\label{steady_state}Steady state}
Taking into account that the steady state of the system is given by the Gaussian state, it is easy to present non-equilibrium steady state in terms of second moments,
We note that $\lgl [x_1, p_1]_+ \rgl = \lgl [x_2, p_2]_+ \rgl = \lgl p_1 p_2 \rgl = 0$. In the high-temperature limit non-zero second moments have the following form,
{\alld
\begin{align} \label{gmss}  
\lgl x_1 ^2 \rgl =& \frac{\gm_1 k T_1(\gm_2 ^2 m^2 \og_0 ^4 - \gm_2 ^2 \kp ^2 + \gm_1 \gm_2 m ^2 \og_0 ^4 + \kp ^2 m^2 \og_0 ^2) +  \gm_2 k T_2 \kp ^2 (m^2\og_0 ^2 + \gm_1 \gm_2)}{(m ^2 \og_0 ^4-\kp ^2) (\gm_1 + \gm_2)(\gm_1 \gm_2 \og_0 ^2 + \kp ^2)} \, ,\non
\\ 
\lgl x_2 ^2 \rgl =& \frac{ \gm_1 k T_1 \kp ^2 (m^2\og_0 ^2 + \gm_1 \gm_2) + \gm_2 k T_2 (\gm_1 ^2 m^2 \og_0 ^4 - \gm_1 ^2 \kp ^2 + \gm_1 \gm_2 m ^2 \og_0 ^4 + \kp ^2 m^2\og_0 ^2)  }{(m ^2 \og_0 ^4-\kp ^2) (\gm_1 + \gm_2)(\gm_1 \gm_2 \og_0 ^2 + \kp ^2)}  \, ,\non
\\ 
\lgl p_1 ^2 \rgl =& m\frac{\gm_1 k T_1 (\gm_2 ^2 \og_0 ^2 + \gm_1 \gm_2 \og_0 ^2 + \kp ^2) + \gm_2 k T_2 \kp ^2 }{(\gm_1 + \gm_2)(\gm_1 \gm_2 \og_0 ^2 + \kp ^2)} \, ,\non
\\  
\lgl x_1 x_2 \rgl =& \frac{ \gm_1 k T_1 \kp + \gm_2 k T_2 \kp }{(\gm_1 + \gm_2)(m ^2 \og_0 ^4-\kp ^2)}\, ,  \non
\\ 
\lgl p_2 ^2 \rgl =& m\frac{\gm_1 k T_1 \kp ^2 + \gm_2 k T_2 (\gm_1 ^2 \og_0 ^2 + \gm_1 \gm_2 \og_0 ^2 + \kp ^2) }{(\gm_1 + \gm_2)(\gm_1 \gm_2 \og_0 ^2 + \kp ^2)} \, , \non
\\   
\lgl x_1 p_2 \rgl = - \lgl x_2 p_1 \rgl =& \frac{\gm_1 k T_2 \gm_2 \kp - \gm_2 k T_2 \gm_1 \kp}{(\gm_1 + \gm_2)(\gm_1 \gm_2 \og_0 ^2 + \kp ^2)}\, . \end{align}}
The non-equilibrium steady-state sec on moments for the weak coupling limit can be obtained from high-temperature ones  by the formal replacement $kT_i\rightarrow \frac{\omega}{2}\coth{\frac{\hb \omega}{2 k T_i}}$.

\bibliographystyle{model1a-num-names}
\bibliography{refs}

\end{document}